\documentclass[10pt,twocolumn]{article}

\usepackage[top=1in,bottom=1in,left=0.75in,right=0.75in,columnsep=0.25in]{geometry}
\usepackage{amsmath, amssymb}
\usepackage{graphicx}
\usepackage{accessibility}
\usepackage{hyperref}
\usepackage{booktabs}
\usepackage{enumitem}
\usepackage{xcolor}
\usepackage{titlesec}
\usepackage{microtype}
\usepackage{wrapfig}

\title{\textbf{LLM-Guided Runtime Parameter Optimization for Energy-Efficient Model Inference}\\}
\author{
Katelyn Crumpacker \\
Virginia Polytechnic and State University \\
\texttt{katelync22@vt.edu}
\and
Dimitrios Nikolopoulos \\
Virginia Polytechnic and State University \\
\texttt{dsn@vt.edu}
}
\date{}

\begin{document}

\maketitle 

\begin{abstract}
Large Language Models (LLMs) have become an integral part of many real-world workflows. However, LLMs consume a lot of energy, which becomes a large concern in the scale of the demand for these tools. As LLMs become integrated into different workflows, different applications have arisen to deal with the challenge of running inference for these tools. This raises another issue of choosing the runtime parameter values for these services in order to minimize the energy consumption. Oftentimes this requires deep knowledge of the application or traditional optimization methods that can take days to find optimal values. In this work, we created a human-in-the-loop flow with LLM-assisted runtime parameter optimization in order to solve this issue. With human-created, specific feedback prompting methods, chat-based LLMs can iteratively find energy-efficient inference parameters faster than traditional search methods. LLMs can also tailor their solutions to different hardware setups and easily take into account other system constraints. The enhanced prompt template was able to converge below the threshold at an average of 3.4 prompts compared to the baseline, which converged in an average of 5.2 prompts, and consistently achieved lower final energy per token. The enhanced prompt template also outperformed Sobol sampling in convergence speed.
\end{abstract}
\section{Introduction}
\indent\textbf{LLM-Assisted Runtime Optimization Framework:} We propose a feedback-driven framework that leverages LLMs to iteratively optimize runtime parameters for energy-efficient LLM inference. \\
\indent\textbf{Energy-Aware Prompt Encoding Strategy:} We design prompt templates using specific prompting techniques that improve LLM performance in the proposed optimization task. \\
\indent\textbf{Cross-Backend Empirical Evaluation:} We evaluate the proposed approach on two inference systems,  vLLM and PyTorch-based deployments, demonstrating generality.\\
\indent\textbf{Reduced Search Overhead:} We show that LLM-guided tuning converges to near-optimal energy per token configurations with fewer evaluations than black-box optimization techniques.

\section{Related Work}
\subsection{Energy-Efficient Model Serving Systems}
Runtime parameter optimization is one method to improve the energy efficiency of LLM inference; there are other techniques that can be used alongside or instead of it. One such solution is ELLIE, which dynamically maps inference phases across heterogeneous hardware for energy reduction. It will make these decisions per prompt and per model, resulting in significant decreases in energy consumption \cite{11113611}. This solution is very effective; however, it can be hardware-dependent or complicated to include in the inference workflow. In a paper by Stojkovic et al., multiple methods, related to techniques like input length and batching, that decrease energy consumption in inference are presented \cite{stojkovic2024towards}.
\subsection{Runtime Parameter Optimization}
There have been other researched methods in parameter optimization.  Bergstra and Bengio provide an argument for random search, as they found it can exceed or match grid search with less computation \cite{10.5555/2188385.2188395}.  Wang et al. propose EcoOptiGen, combining Bayesian optimization with cost-aware pruning for LLM inference \cite{pmlr-v224-wang23b}. Zhang et al. use an LLM as the optimizing tool for machine learning models, achieving competitive performance with Bayesian optimization \cite{zhang2024usinglargelanguagemodels}. However, their work focuses on optimizing machine learning model hyperparameters rather than runtime parameters of inference systems.
\section{Design}
The goal of our design is to evaluate whether LLMs can guide runtime parameter optimization for energy-efficient LLM inference under realistic deployment constraints.  We use the LLM as an adaptive decision agent, iteratively proposing configurations based on observed system feedback rather than relying on exhaustive search. To validate the LLM’s ability to optimize runtime parameters for inference engines, we selected vLLM to run on a single 16GB GPU for text-based prompts. vLLM is memory-efficient, which allows for a larger search space. For further validation, we used PyTorch for image and text workloads. PyTorch and vLLM exhibit complex cross-layer interactions between runtime scheduling, memory usage, and computation that motivate an LLM-guided black-box optimization strategy. Our search space includes parameters that fall into four categories: (i) memory utilization, (ii) batch and execution control, (iii) precision and computation, and (iv) hardware constraints. We selected these because they influence energy consumption through their effects on GPU occupancy, device utilization, computational cost, and power limits. Although evaluated on vLLM and PyTorch, we designed the framework to remain runtime-agnostic and generalizable to other inference systems.  
\subsection{LLM-Guided System-Level Parameter Optimization}
The system involves using an LLM to improve inference of other LLMs. LLMs maintain context across iterations, allowing them to incorporate prior information when proposing new configurations \cite{wu2025humanmemoryaimemory}. This differs from search methods like Sobol sampling. We chose Sobol sampling as a baseline comparison because it deterministically finds configurations that cover the search space uniformly. We chose Claude as the online LLM because it is freely accessible via a web interface. Claude offers a project feature for chats to have isolated memory \cite{anthropic2026claude-chat-search-memory}.

\begin{figure}[h]
    \centering
    \includegraphics[width=\linewidth,\alt{showing iterative energy optimization using LLM feedback}]{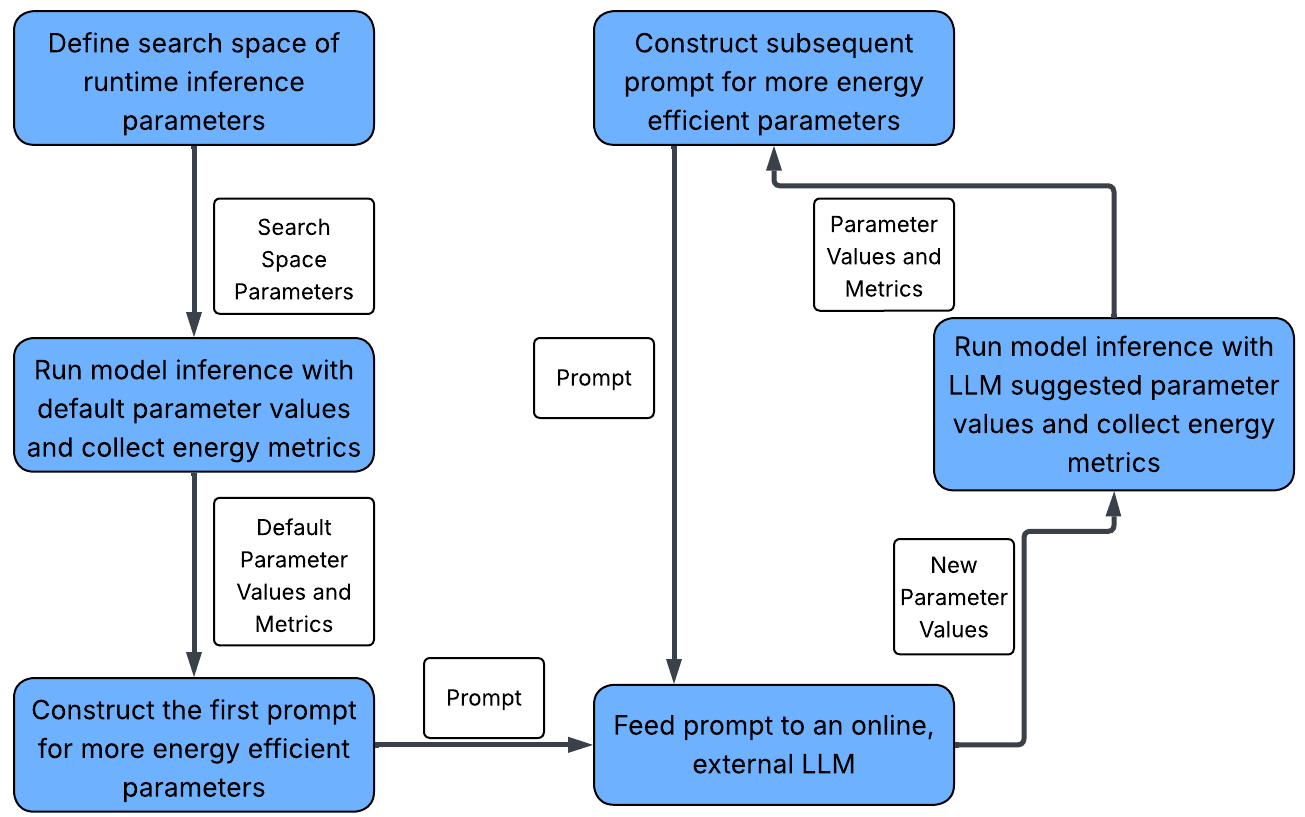}
    \caption{Human-in-the-loop workflow for optimizing runtime inference parameters using an external LLM. Energy metrics from each inference run are used in subsequent prompts to guide parameter exploration.}
    \label{fig:Human-in-the-Loop}
\end{figure}

\subsection{Prompting Construction and Feedback Encoding}
We developed a workflow (Figure \ref{fig:Human-in-the-Loop}) that includes a search space, human-in-the-loop prompting of LLMs, and energy metric collection. We incorporated human oversight to monitor proposals and enforce deployment constraints \cite{10.1007/978-3-031-14135-5_7}. We constructed prompts based on energy metrics of the previous configuration to guide the LLM toward more energy-efficient parameter configurations. We tested prompt techniques to determine which combinations helped the LLM produce the most effective results. We explored repetition, along with adding background information in the first prompt, which the LLM can use throughout iteration.  We also tested structured versus unstructured prompts \cite{wen2025thinkpatterns21ksystematicstudyimpact}. For the structured prompt strategy, we used search-process constraints in guiding the LLM and keeping its focus on a single task (enhanced prompt template). We also experimented with unstructured prompts that leave the LLM open to focus on more multi-task optimizations (baseline prompt template).

\section{Implementation}
\begin{figure}[h]
    \centering
    \includegraphics[width=\linewidth,\alt{Diagram of a pipeline where user input is processed by a language model, evaluated using energy metrics, and iteratively refined through feedback to improve efficiency.}]{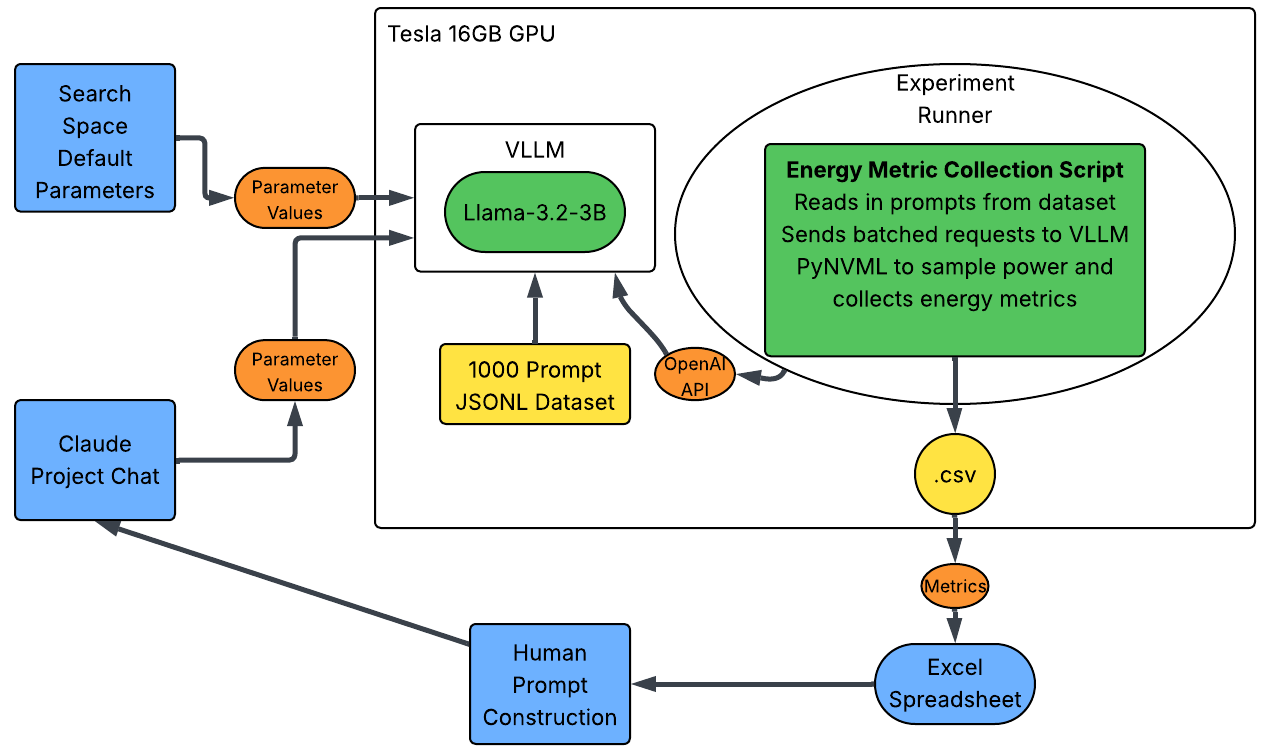}
    \caption{System architecture for measuring energy consumption during LLM inference. An experiment runner executes prompts through vLLM while a PyNVML-based script samples GPU power to compute energy metrics. Parameter configurations are provided from the defined search space or suggested by an external LLM.}
    \label{fig:System-Diagram}
\end{figure}

\subsection{System Setup}
Figure \ref{fig:System-Diagram} shows the end-to-end system architecture used in the experiments to see the effect prompts have on the LLM’s ability to generate candidate runtime parameters.
\subsubsection{Hardware and Software Stack}
We collected data on a server with a single NVIDIA Tesla V100 PCIe GPU with 16 GB of device memory. We set the \texttt{gpu-memory-utilization} flag to 0.55 to reduce excessive OOM errors. vLLM serves the decoder-only LLM used for inference. The client concurrently issues inference requests, while batching decisions are handled entirely by vLLM at the token level. The system runs on top of PyTorch with CUDA-enabled kernels. An experiment runner orchestrates the execution loop. This runner is responsible for loading prompt data, configuring inference parameters, issuing batched inference requests to vLLM, and logging measurements in a .csv file.
\subsubsection{Model and Dataset}
We evaluate Llama-3.2-3B served through vLLM. Inference requests are generated from a fixed dataset of 1,000 prompts. We reuse the dataset across all experimental runs to ensure comparability across different parameter configurations. 
\subsubsection{Energy Measurement}
The recorded metrics are total energy consumption in joules (J), wall time in seconds (s), total tokens, energy per token in J/token, and throughput in tokens/s. We measured the GPU energy via NVIDIA’s Management Library (NVML). We sample GPU power using NVML’s \texttt{nvmlDeviceGetPowerUsage} API. Energy measurement begins before issuing the first inference request and ends after the final request completes, capturing the GPU energy cost of batched inference, including request scheduling and queuing effects. To study the impact of power constraints, a GPU power limit is optionally enforced when specified by the LLM, using NVML’s power management interface. 
\subsubsection{Parameter Sources and Control Flow}
Parameters originate from two sources: (1) a default search space specification defining valid parameter ranges, and (2) parameter values generated by an external controller. For the latter, a project-specific Claude chat produces parameter configurations based on prior experiment outcomes. We inject these values directly into the experiment runner, allowing the system to evaluate each configuration. Human intervention is limited to prompt construction and experiment supervision. We do not perform manual tuning during execution; all parameter changes occur between experiment runs.
\subsubsection{Sobol Sampling}
To establish a comparison baseline, we implement a Sobol-sample–based search over the same parameter space used by the LLM-driven controller. The baseline evaluation consists of 30 Sobol-sampled configurations for the vLLM search space, which are executed sequentially using the same parameter space and evaluated under the same conditions. Both the Sobol baseline and the LLM-guided approach use the same execution environment, parameter domains, and metric methodology.
\subsection{Parameter Space}
 The search space consists of four categories: token budget and context parameters (\texttt{max-num-batched-tokens} [256–4096] and \texttt{max-model-len} [256–4096]), memory system parameters (\texttt{block-size} [16, 32]), concurrency parameters (\texttt{max-num-sequences} [32–256] and \texttt{max-num-partial-prefills} [1–10]), and a hardware-level parameter (\texttt{power-limit} [100–250 W]). 
\begin{figure}[h]
    \centering
    \includegraphics[width=\linewidth,\alt{Prompt template for first and subsequent prompts that is designed to guide a language model in selecting energy-efficient inference parameters that is enhanced using structured prompting strategies.}]{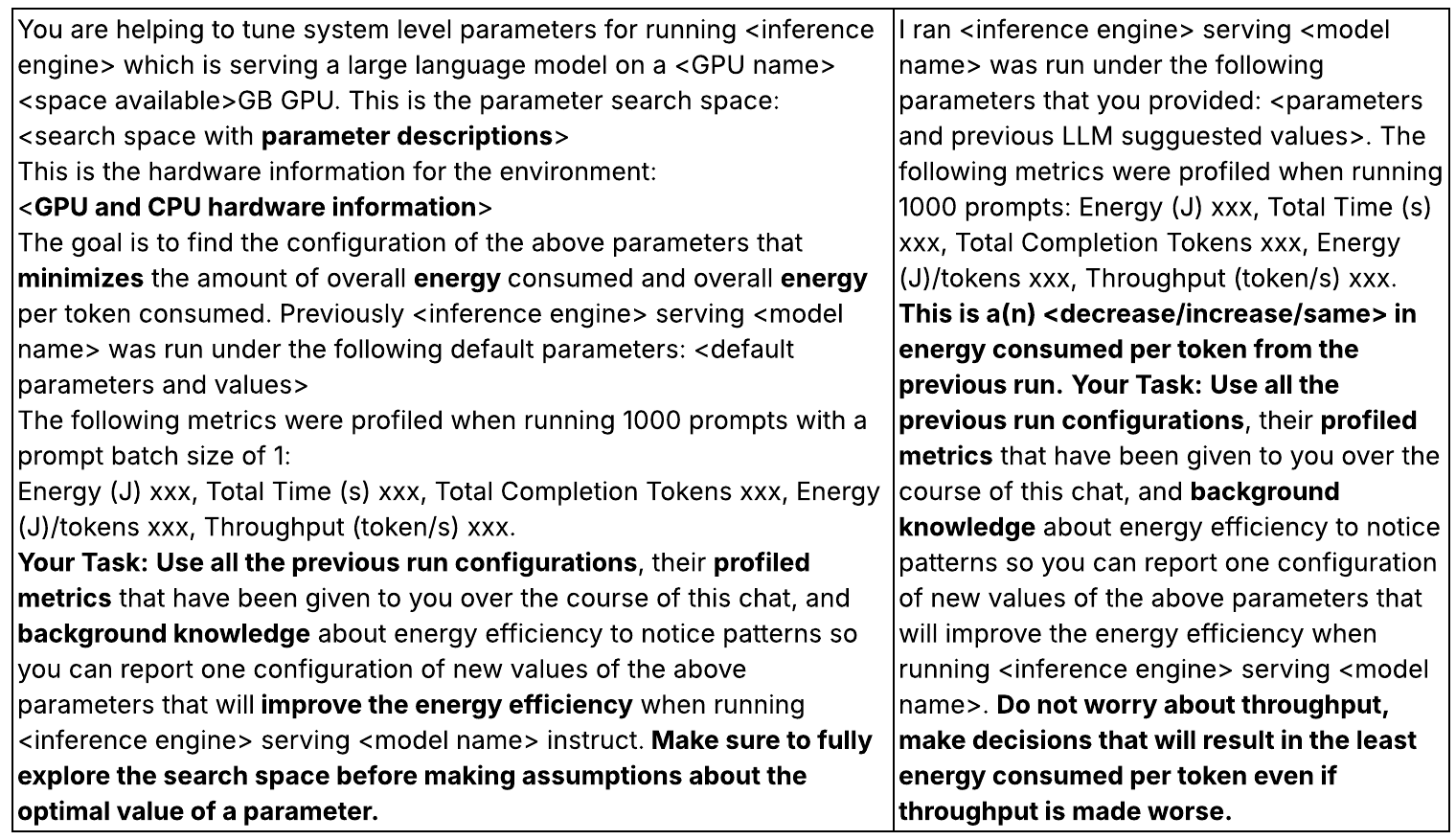}
    \caption{Template for the enhanced prompts including the first prompt (on the left) and subsequent prompts (on the right). Prompting techniques discussed in section 4.3 are highlighted in bold.}
    \label{fig:Enhanced-Prompt}
\end{figure}

\begin{figure}[h]
    \centering
    \includegraphics[width=\linewidth,\alt{Prompt template for first and subsequent prompts that is designed to guide a language model in selecting energy-efficient inference parameters.}]{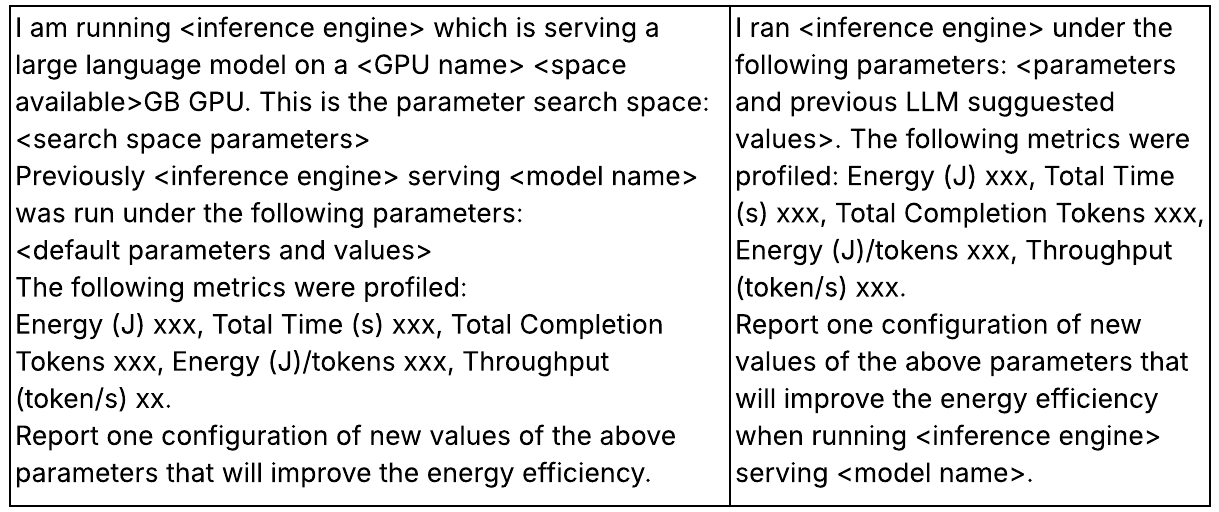}
    \caption{Template for the baseline prompts including the first prompt (on the left) and subsequent prompts (on the right).}
    \label{fig:Baseline-Prompt}
\end{figure}
\subsection{LLM Integration and Control Plane}
We integrate Claude into this experiment prompting it to generate the energy-efficient runtime parameters that vLLM will use to serve the Llama model. We separate the prompt templates into enhanced and baseline categories. Each category contains a first prompt template and the subsequent prompt template. The first prompt template describes the task, provides the search space, gives the default parameter configuration and metrics, and gets the LLM-suggested parameter configuration. Subsequent prompts provide the previous runtime parameters and the metrics. The enhanced prompt (Figure \ref{fig:Enhanced-Prompt}) uses the structured approach with repetition, background, search-process guidance, and task framing. The background we provide is the hardware information and parameter descriptions. The baseline prompt (Figure \ref{fig:Baseline-Prompt}) provides the minimum information the LLM needs, in a more unstructured approach. Invalid configurations trigger a follow-up prompt providing the error, restating the goal, and asking for a new configuration, without halting the iterative loop. 
\subsection{Multimodal PyTorch Inference Setup}
To evaluate whether the LLM-guided parameter optimization strategy generalizes beyond vLLM, we conduct additional experiments using the multimodal model Qwen3-VL-4B-Instruct executed directly in PyTorch. In Figure \ref{fig:System-Diagram}, the system architecture remains unchanged, with vLLM replaced by PyTorch, the text dataset replaced by a 500 image multimodal dataset, and we we gathered additional energy metrics (energy per image embedding). Each image is paired with randomly sampled questions to form image-text prompts.  The search space includes: token budget and context parameters (\texttt{pixel-precision} [fp16, fp32]), memory system parameters (\texttt{cuda-memory-fraction} [0.1–1.0] (step 0.1)), concurrency parameters (\texttt{batch-size} [1-16]), and a hardware-level parameter (\texttt{power-limit} [100–250 W]). Sobol sampling is also performed on this search space using 64 configurations as opposed to the 30 configurations for vLLM (image inference ran faster so more configurations could be completed in the same timeline).

\section{Evaluation}
We evaluated Claude combined with the enhanced and baseline prompting strategies on their ability to generate energy-efficient parameters for vLLM inference, as well as PyTorch inference. We collected data by running the prompt templates, through the loop shown in Figure \ref{fig:Human-in-the-Loop}. We repeated this loop five times for both the enhanced and baseline prompt templates.

\begin{figure}[h]
    \centering
    \includegraphics[width=\linewidth, \alt{Over iterations, enhanced prompts reach lower energy per token faster with vLLM serving LLama than baseline prompts.}]{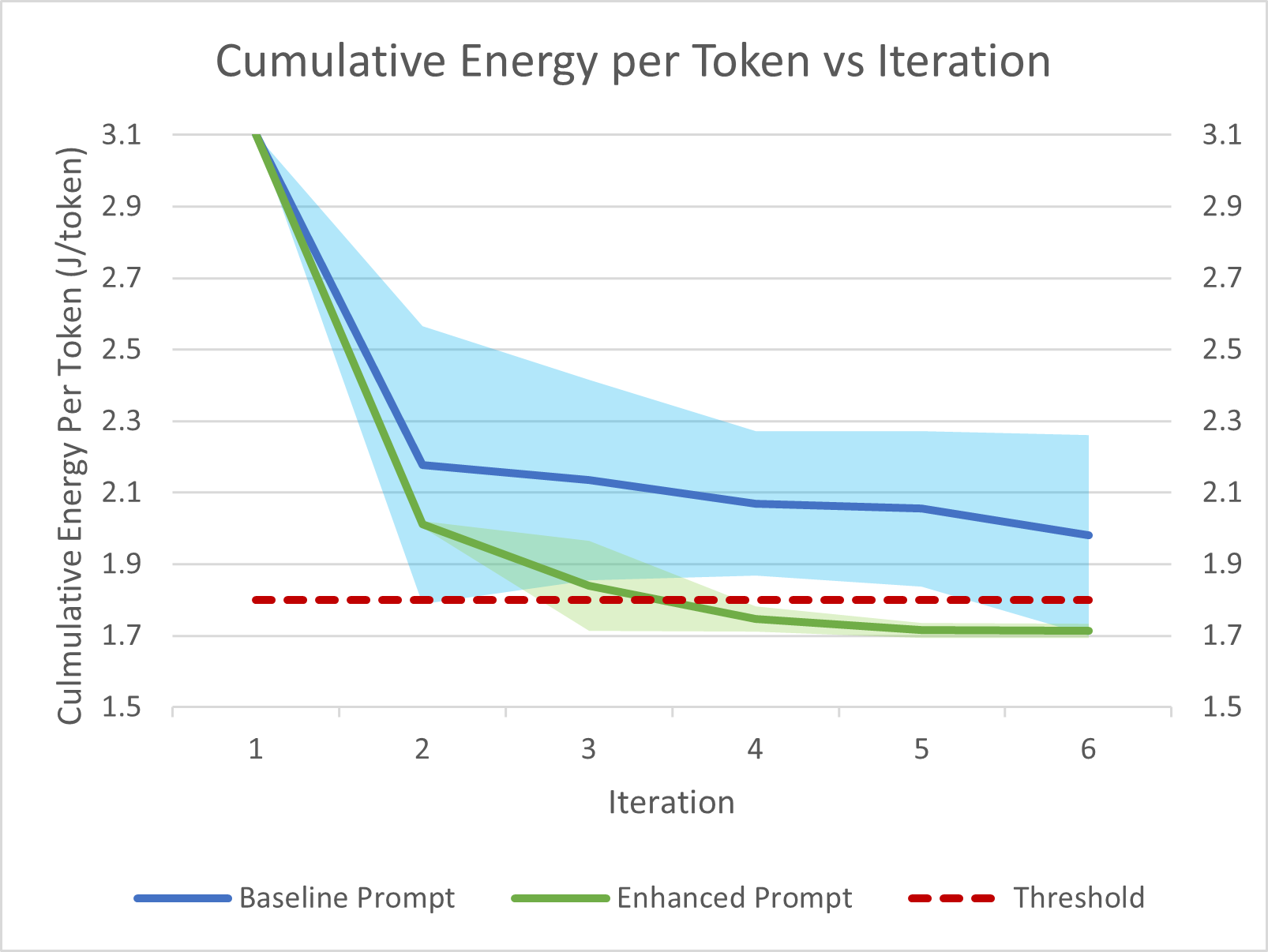}
    \caption{Cumulative energy per token over iterations for baseline and enhanced prompts across five runs. Shaded areas represent ±1 standard deviation. The enhanced prompt reaches the convergence threshold (1.80 J/token) earlier and maintains lower energy values with smaller variance, demonstrating faster and more consistent convergence toward energy-efficient configurations.}
    \label{fig:Cumulative-Graph}
\end{figure}
\subsection{Optimization Effectiveness Across Prompting Strategies}
To evaluate the effectiveness of the enhanced prompt template compared to the baseline, we analyzed convergence speed and final energy efficiency for Llama-3.2-3B inference via vLLM on a single 16GB GPU. We define a convergence threshold of 1.80 J/token for energy per token. This represents approximately a 42\% reduction from the default configuration value of 3.1 J/token.  From a practical standpoint, further reductions to far below this threshold provide diminishing returns due to hardware and memory constraints, making 1.80 J/token a reasonable target for energy-efficient inference. The 1.80 J/token threshold was selected post-hoc as a reference point near the empirically observed minimum, serving as a convergence marker rather than a performance target.
\subsubsection{Convergence Speed Analysis} 
For each of five independent runs, we measured the number of iterations required to reach the threshold. Runs where the baseline did not reach the threshold within the maximum allowed iterations were assigned the maximum iteration count (3 out of 5 runs required this conservative estimate for the baseline prompts). Enhanced prompts consistently reached the threshold in fewer iterations than baseline (Mean $\pm$ SD: 3.4 $\pm$ 0.55 vs 5.2 $\pm$ 0.84 iterations). A paired t-test confirms that this difference is statistically significant (t(4) = 4.81, p = 0.0086, two-tailed), with a large effect size (Cohen's d $\approx$ 2.55). Figure \ref{fig:Cumulative-Graph} further demonstrates that the enhanced prompt accelerates convergence toward energy-efficient configurations and has less variance in energy per token values across runs.
\begin{table}
\centering
\resizebox{\linewidth}{!}{%
\begin{tabular}{|c|c|c|}
\hline
\textbf{Experiment Run} & \textbf{Baseline Min Energy per Token} & \textbf{Enhanced Min Energy per Token} \\
\hline
1 & 1.75724 & 1.68337\\
2 & 1.99868 & 1.71955 \\
3 & 2.42611 & 1.74051 \\
4 & 1.72301 & 1.71165 \\
5 & 1.99748 & 1.71461 \\
\hline
\end{tabular}%
}
\caption{Minimum energy per token values across experiment runs for baseline and enhanced prompt templates.}
\label{tab:energy_per_token}
\end{table}
\subsubsection{Final Energy per Token} While convergence speed is critical, final energy efficiency is also important. Across all runs, the enhanced prompt consistently achieved lower cumulative minimum energy per token than baseline. Table 1 summarizes the final minimum energy per token per run and it can be seen that for all five experiments the enhanced prompt reaches a lower energy per token value than the baseline prompt.
\subsection{Sobol Sampling Comparison}
Executing the Sobol-based experiment runner required approximately 24 hours to evaluate 30 parameter configurations, each with 1000-prompt inference. Among these, Sobol configuration six achieved the lowest energy per token value of 1.72 J/token, comparable to the best result obtained using the enhanced LLM prompting strategy. However, identifying this configuration required full evaluation of all 30 Sobol samples. The LLM enhanced prompt combination reached approximately the same energy per token value in an average of 3.4 prompts. For PyTorch inference the Sobol-based experiment runner required approximately 6 hours to evaluate 64 parameter configurations. The majority of configurations generated by Sobol sampling (55 out of the 64) resulted in an OOM error (because pixel\_precision was set to fp32). Sobol Sampling has an energy per token value in run 3 of 0.141 J/token which is comparable to the best result obtained using the enhanced LLM prompting strategy which is around 0.160 J/token and took an average of five runs to achieve. However, Sobol sampling never reaches the best result in terms of energy consumed per image embedding achieved by the enhanced prompts. On run 6, Sobol sampling's energy consumed per image is 82.5 J/image embedding. This is compared to 27.5 J/image embedding that the enhanced prompting strategy achieves on average after five prompts. 
\begin{figure}[h]
    \centering
    \includegraphics[width=\linewidth,\alt{Over iterations, enhanced prompts reach lower energy per token faster than baseline prompts for PyTorch serving Qwen.}]{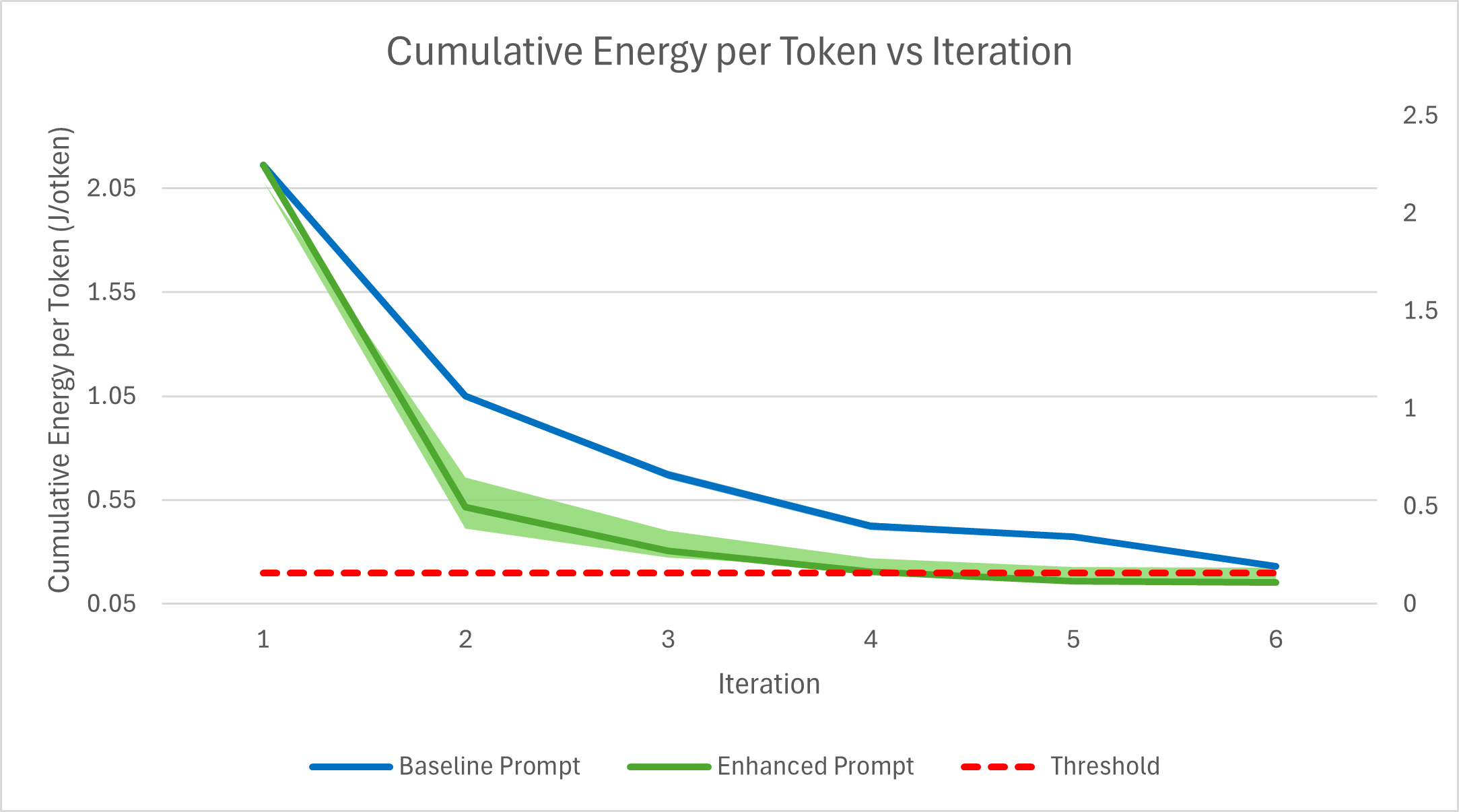}
    \caption{Cumulative energy per token over iterations for baseline and enhanced prompts across five runs when running Qwen with PyTorch. Shaded areas represent ±1 standard deviation (shaded for the baseline is too small to be shown over the line). The enhanced prompt reaches the convergence threshold (0.20 J/token) earlier and maintains lower energy values with smaller variance, demonstrating faster and more consistent convergence toward energy-efficient configurations.}
    \label{fig:Cumulative-Image}
\end{figure}
\begin{figure}[h]
    \centering
    \includegraphics[width=\linewidth, \alt{Over iterations, enhanced prompts reach lower energy per image faster than baseline prompts for PyTorch serving Qwen.}]{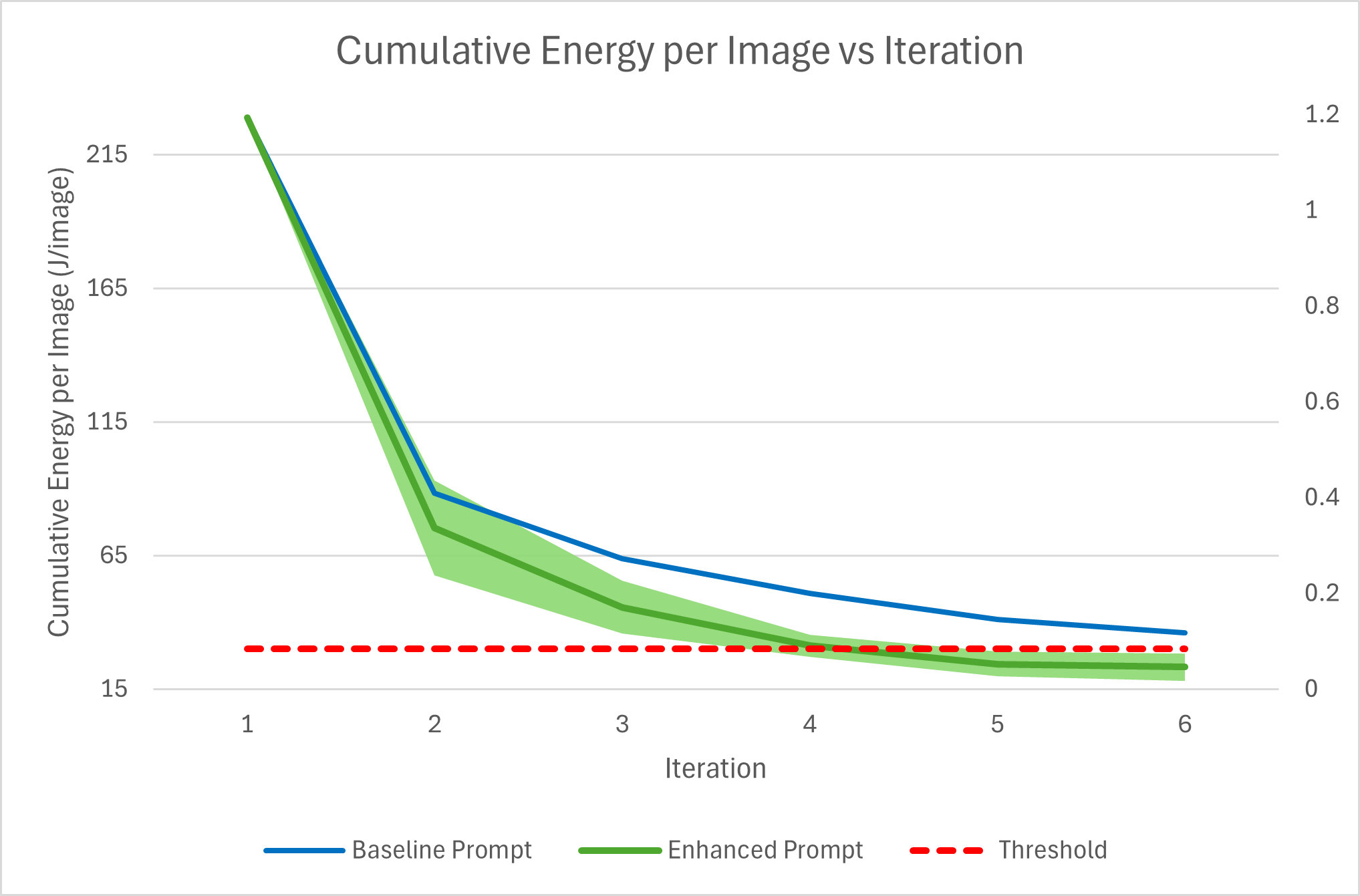}
    \caption{Cumulative energy per image over iterations for baseline and enhanced prompts across five runs when running Qwen with PyTorch. Shaded areas represent ±1 standard deviation (shaded for the baseline is too small to be shown over the line). The enhanced prompt reaches the convergence threshold (30 J/image) earlier and maintains lower energy values with smaller variance, demonstrating faster and more consistent convergence toward energy-efficient configurations.}
    \label{fig:Cumulative-Image2}
\end{figure}
\subsection{Multimodal Generalization}
To confirm the effectiveness of the enhanced prompt template compared to the baseline, we analyzed convergence speed and final energy efficiency for Qwen3-VL-4B-Instruct inference via PyTorch on a single 16GB GPU. We define a convergence threshold of 0.20 J/token for energy per token and 30 J/image for energy per image. This represents approximately a 90.7\% reduction from the default configuration value of 2.16 J/token and approximately an 87\% reduction from the default configuration value of 228.9 J/image. From a practical standpoint, further reductions to far below this threshold provide diminishing returns due to hardware and memory constraints, making 0.20 J/token and 30 J/image reasonable targets for energy-efficient inference. The 0.20 J/token and 30 J/image thresholds were selected post-hoc as a reference point near the empirically observed minimum, serving as a convergence marker rather than a performance target.
\subsubsection{Convergence Speed Analysis} 
For each of five independent runs, we measured the number of iterations required to reach the threshold. Runs where the baseline did not reach the threshold within the maximum allowed iterations were assigned the maximum iteration count (all 5 runs required this conservative estimate for the baseline prompts). Enhanced prompts consistently reached the energy per token threshold, as well as the energy per image threshold, in fewer iterations than baseline. The enhanced prompts reached the energy per token threshold in an average of 4.2 iterations (Mean $\pm$ SD: 4.2 $\pm$ 0.84). The baseline prompts (with the conservative estimate) reached the energy per token threshold in an average of 6 iterations (Mean $\pm$ SD: 6 $\pm$ 0). A paired t-test showed that this difference was statistically significant (t(4) = 4.00, p = 0.016, two-tailed), with a large effect size (Cohen’s d $\approx$ 1.79). The average iterations taken to reach the threshold was the exact same for the energy per image threshold across enhanced and baseline prompts. Figure \ref{fig:Cumulative-Image} further demonstrates that the enhanced prompt accelerates convergence toward energy-efficient configurations in regards to energy per token. Differing from the vLLM experiments, the enhanced prompt has more variance across iterations than baseline prompts. However, all experimental runs experienced the same trend of continuing to decrease the energy per token. Figure \ref{fig:Cumulative-Image2} shows that there are similar results in regards to energy per image. The enhanced prompt accelerates convergence and experiences the same change in variance across iterations. 
\subsubsection{Final Energy per Token} While convergence speed is critical, final energy efficiency is also important. Across all runs, the enhanced prompt consistently achieved lower cumulative minimum energy per token values and energy per image values than the baseline. In all experimental runs, the baseline prompts were not able to reach the thresholds within the limit of 5 prompts. 
\subsection{Optimization Overhead}
To quantify the energy overhead of online LLM-guided optimization, we conservatively estimate the energy cost per external prompt as 3 Wh. We base this value on previously reported measurements of earlier Claude-class models \cite{jegham2025hungryaibenchmarkingenergy}. The enhanced configuration requires an average of 3.4 optimization prompts. For conservatism and clarity, we round this value up to 4 prompts. Assuming an upper-bound estimate of 3~Wh per external LLM prompt, the average total Wh consumed is 12 which is 43,200 J. The average energy consumed when testing out the configurations provided by the 4 LLM prompts is 2,162,832 J making the total energy consumption 2,206,032 J. The optimized configuration (1.80~J/token threshold) saves an average of 390{,}401~J per 1000-prompt workload relative to the default configuration. The break-even point occurs when cumulative energy savings equal the optimization overhead and is 5.65 workloads. This indicates that the optimization overhead is negligible after six 1000-prompt workloads.
\begin{figure}[h]
    \centering
\includegraphics[width=\linewidth, \alt{LLM-optimized configurations achieve higher throughput and lower energy per token than the default configuration, moving the system toward a more Pareto-efficient operating region.}]{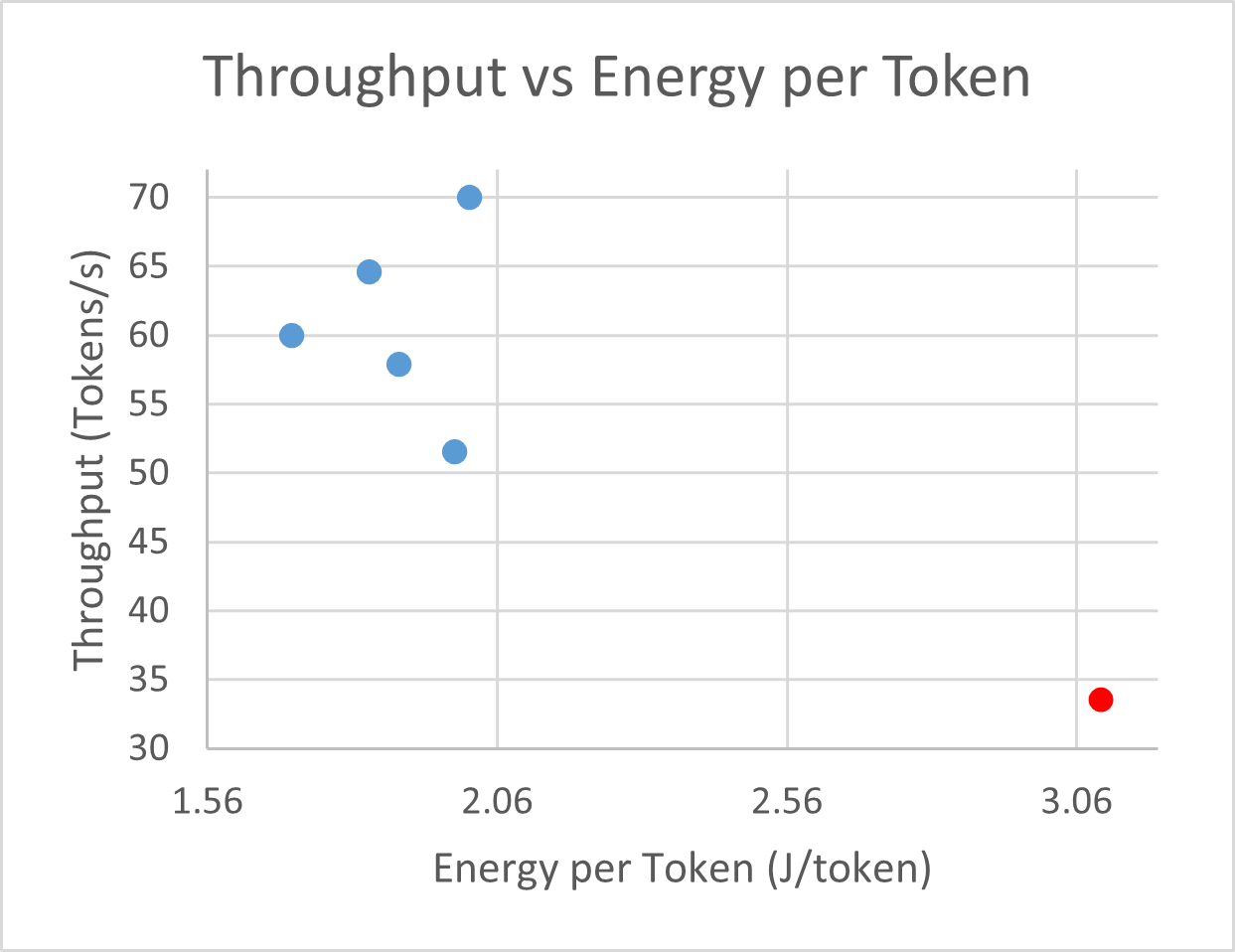}
    \caption{Throughput versus energy per token for evaluated configurations. The default configuration (red) consumes significantly more energy while achieving lower throughput. LLM-optimized configurations move the system toward a more Pareto-efficient operating region with both lower energy consumption and higher throughput.}
    \label{fig:PeformanceVEnergy}
\end{figure}
\subsection{Performance vs Energy Efficiency Tradeoffs}
In HPC, improving energy efficiency often comes at the expense of reduced performance. To evaluate whether this tradeoff occurs in our setting, we use inference throughput (tokens/s) as the primary performance metric (Figure \ref{fig:PeformanceVEnergy}).  Despite optimizing exclusively for energy per token (J/token), the optimized configurations increase average throughput from 33.5 tokens/s under the default configuration to 59 tokens/s, representing a 76\% improvement. From an HPC perspective, the optimized configurations move the system toward a more Pareto-efficient operating region, simultaneously improving throughput.

\section{Discussion}
In HPC environments, the results show that LLMs can make metric-aware decisions without exhaustive search, aligning with emerging work on AI-assisted system optimization. In this context, scalability refers to optimization problems with large parameter spaces and tighter hardware constraints, which create complex interactions between runtime parameters. Uniform exploration strategies require a large number of evaluations to maintain adequate coverage across these high-dimensional spaces, and each evaluation incurs the cost of GPU execution. Our results suggest that LLMs combined with prompt engineering can handle this without increasing evaluations. From a systems perspective, reducing the number of required evaluations improves the scalability of the optimization process by lowering the runtime and computational overhead associated with configuration search. Because our experiments were conducted in a single-node, single-GPU environment, our results do not directly generalize to multi-node inference systems, which introduce additional parameters and system-level interactions affecting energy consumption, but there is a lot of potential for LLMs to handle the increased search space and complex deployment information. Inference at this scale also makes exhaustive search methods take longer and incur more overhead to get an optimal parameter configuration. Variations in LLM behavior, prompting strategy, or stochastic outputs can affect reproducibility. However, with the prompting templates in this work, this issue can be minimized.

\section{Conclusion}
This work demonstrates that LLM-guided optimization, combined with prompt engineering, can reduce the exploration cost, and improve energy efficiency in HPC-relevant workloads. Future work should evaluate LLMs in multi-node inference deployments. Although throughput improved in our experiments, it was not maximized as this study focused on minimizing energy per token. However, future research could explore LLMs' ability to find the best configuration that maximizes throughput and minimizes energy per token. Overall, these results highlight the potential of LLMs to enable adaptive runtime optimization for energy efficiency in HPC environments.

\bibliographystyle{plain}
\bibliography{references}
\end{document}